\documentclass[preprint2]{emulateapj-rtx4}

\usepackage{txfonts} 
\usepackage{epsfig} 
\usepackage{graphicx}
\usepackage{natbib} 
\usepackage{xspace}
\usepackage{color}
\usepackage[colorlinks,urlcolor=blue,citecolor=blue,linkcolor=blue]{hyperref} 

\newcommand{\kms}{{\km\second^{-1}}}
\newcommand{\solarmass}{{\,{\textnormal{M}}_\odot}}

\newcommand{\km}{\,{\textnormal{km}}}
\newcommand{\second}{\,{\textnormal{s}}}

\newcommand{\dechms}[4]{$#1^{\rm h}#2^{\rm m}#3\mbox{$^{\rm s}\mskip-7.6mu.\,$}#4$}

\newcommand{\decdms}[4]{$-#1^{\circ}#2'#3\mbox{$''\mskip-7.6mu.\,$}#4$}

\newcommand{\mo}    {$M_{\sun}$}

\begin{document}

\title{Origin and kinematics of the eruptive flow from {\it XZ Tau} revealed by ALMA}

\shortauthors{Zapata, et al.}

\author{Luis A. Zapata\altaffilmark{1}, Roberto Galv\'an-Madrid\altaffilmark{1}, Carlos Carrasco-Gonz\'alez\altaffilmark{1}, Salvador Curiel\altaffilmark{2}, 
Aina Palau\altaffilmark{1}, \\ Luis F. Rodr\'\i guez\altaffilmark{1}, Stan E. Kurtz\altaffilmark{1},  Daniel Tafoya\altaffilmark{1}, and Laurent Loinard\altaffilmark{1}} 

\altaffiltext{1}{Centro de Radioastronom\'\i a y Astrof\'\i sica, UNAM, Apdo. Postal 3-72
  (Xangari), 58089 Morelia, Michoac\'an, M\'exico}
\altaffiltext{2}{Instituto de Astronom\'\i a, Universidad Nacional Aut\'onoma de M\'exico, Ap. 70-264, 04510 DF, M\'exico} 

\begin{abstract} 
We present high angular resolution ($\sim$0.94$''$) $^{12}$CO(1-0) {\it Atacama Large Millimeter/Submillimeter Array} (ALMA) 
observations obtained during the 2014 long baseline campaign from the eruptive bipolar flow from the multiple {\it XZ Tau} stellar system discovered 
by the {\it Hubble Space Telescope} (HST). These observations reveal, for the first time, the kinematics of the molecular flow.  
The kinematics of the different ejections close to {\it XZ Tau} reveal a rotating and expanding structure 
with a southeast-northwest velocity gradient.  The youngest eruptive bubbles unveiled in the optical HST images are inside of this molecular 
expanding structure.  Additionally, we report a very compact and collimated bipolar outflow emanating 
from {\it XZ Tau A}, which indicates that the eruptive outflow is indeed originating from this object. 
The mass (3 $\times$ 10$^{-7}$ $\solarmass$) and energetics (E$_{kin}$ $=$ 3 $\times$ 10$^{37}$ ergs) 
for the collimated outflow are comparable with those found in molecular outflows associated with young brown dwarfs. 
\end{abstract}


\section{Introduction}

XZ Tau is considered a classical T Tauri star (classified as a class I/II), which is composed of a binary system,  XZ Tau (A/B).
The binary system is separated by approximately 0.3$''$ (at a P.A. of 154$^\circ$), or 42 AU at a distance of  
150 pc \citep{tor2009}. The relative flux-ratio between the two components fluctuates at optical and radio wavelengths, probably because of
non-thermal (gyro)synchrotron emission associated with magnetic activity \citep{kri2008}.
XZ Tau is exciting an optical outflow (HH 151) with a P.A. $=$ 15$^\circ$, and which can be traced over about 10$''$
on either side of the star as revealed by the studies of \citet{mun1990}.

Contrary to the long and well-collimated jet emanating from the nearby T Tauri star HL Tau \citep{kri2008,lum2014}, XZ Tau is driving
an outflow with a morphology reminiscent of numerous concentric explosions or hot bubbles,   
as revealed by multiple observations carried out with the {\it Hubble Space Telescope} (HST) \citep{kri1999,kri2008}. 
There are only two reported additional cases where this morphology ({\it i.e.} bubble-like) is present: the SVS 13 outflow \citep{hod2014}
and IRAS16293$-$2422(B) \citep{loi2013}. The nature of these bubble-like outflows (including that of XZ Tau) is still under debate. 

The expansion (tangential, or in the plane of the sky) velocities of the elongated optical bubbles in the flow from XZ Tau are about 70-200 $\kms$ 
\citep{kri2008}, while the radial velocities (in the line-of-sight) are about 100-300 km s$^{-1}$ 
obtained from optical images of forbidden lines of nitrogen [NII $\lambda$6583] or sulfur [SII $\lambda$6716] \citep{hir1997}.  
Comparing the radial velocities and proper motions of the outflow, it has been suggested that the bubbles are mostly in the plane
of the sky with an angle of about 30$^\circ$ with respect to this plane \citep{kri2008}.  
Multiple images from the HST have revealed that the eruptive outflow is episodic, probably excited by  
velocity pulses in the outflow of XZ Tau A \citep{kri2008}.

\citet{kri2008} suggested that bubbles are ejected from XZ Tau A because this 
source is better aligned to the flow and XZ Tau B seems very faint \citep{hir1997}.
\citet{kri1999,kri2008} proposed that possibly the periastron passage of 
XZ Tau B might produce the eruptive outflow.
However, the short dynamical ages for the bubbles cannot be explained 
by the very slow motions of the binary system XZ Tau (A/B) in the plane of the sky \citep{kri2008,car2009,for2014}. 
\citet{car2009} reported the detection of a close radio binary system in XZ Tau A with a separation of only 0.09$''$ (13 AU), 
and probably in an elliptical orbit. They suggested that the periastron passage of one component, XZ Tau C 
(named S2 by Carrasco-Gonz\'alez et al. 2009), of the discovered close radio binary system could much better 
explain the eruptions in the flow from XZ Tau. 

It is speculated that the mass of the binary system (A/C) is around 1 \mo, with a period for the system 
of only 40 yr \citep{car2009}.  Therefore, the A/C system was at its previous periastron in the 1980s, and will be again at this position
around 2020. However, the data presented in  \citet{car2009} are not sufficient to confirm or reject this hypothesis, so  
this explanation for the bubble-like outflow generation was tentative.
In addition, \citet{for2014} did not detect the XZ Tau C component previously reported by  \citet{car2009}.
More observations of XZ Tau A/C at around 2020 would be required to strongly test the periastron passage 
model for bubble-like outflow generation \citep{for2014}.

 In this {\it Letter}, we present new, high angular resolution $^{12}$CO(1-0) line observations obtained with the 
{\it Atacama Large Millimeter/Submillimeter Array} (ALMA) toward the eruptive outflow from XZ Tau. Our analysis reveals 
the kinematics of the eruptive flow, as well as a very compact bipolar jet emanating from XZ Tau A.
This compact molecular outflow is at the base of the optical eruptive flow.

\begin{figure*}[!]
\begin{center}
\includegraphics[scale=0.39]{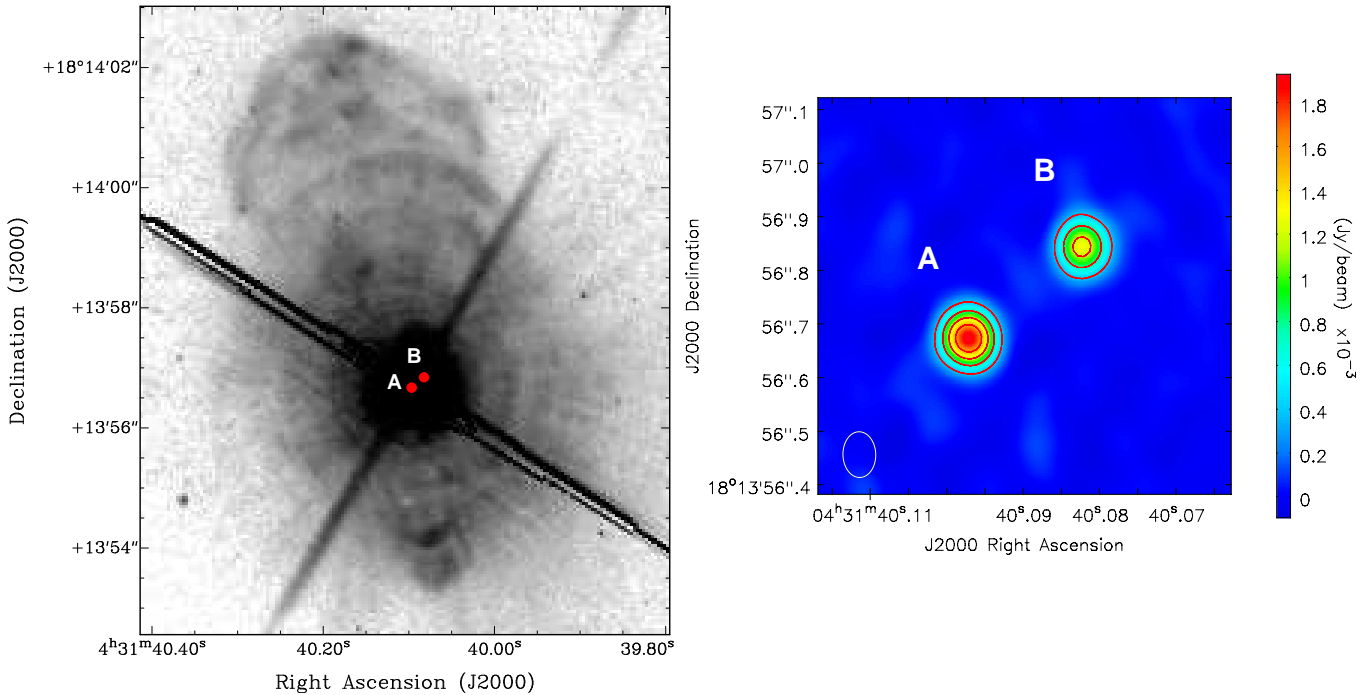}
\caption{ Left: HST/ACS (narrow band FR656N, grey scale) image overlaid with the positions of the 3 mm continuum sources  
from the eruptive flow in the XZ Tau region. Right:  3 mm (contour and color image) ALMA images from the 
XZ Tau A/B system.  The continuum ALMA images were taken from \citet{par2015}. 
The red contours are 0.386, 0.772, 1.16 and 1.54 mJy beam$^{-1}$. The synthesized beam of the ALMA  
continuum image is shown in the lower-left corner of the right image.
The optical image has an offset of about 0.4$''$ to the south in order to correct the proper motions since the epoch of the HST observations, 
and the World Coordinate System (WCS) uncertainties for the HST images.}
\label{fig1}
\end{center}
\end{figure*}

\section{Observations}

The CO observations were made with 32-35 antennas \citep{par2015,pare2015} of ALMA in 2014 October and November,
 during the ALMA science verification data program for the long baseline campaign. 
The purpose of this campaign was to verify and demonstrate ALMA's performance 
on baseline lengths larger than 10 km. 
These observations only included antennas with diameters of 12 meters.  The
independent baselines ranged in projected length from 12 to 15,240 m.  
The phase center of these millimeter observations was at $\alpha_{J2000.0}$ = 
\dechms{04}{31}{38}{426}, $\delta_{J2000.0}$ = \decdms{18}{13}{57}{04}, the position of the nearby HL Tau. 
The primary beam of ALMA at 115 GHz has a FWHM of $\sim$55 arcsec, so it included XZ Tau within 
the primary field of view. As XZ Tau was at the eastern side of our FWHM, we corrected the final continuum image  
for the primary beam attenuation.   

The ALMA digital correlator was configured to observe the $^{12}$CO(1-0) line at a rest 
frequency of 115.27120 GHz. The spectral window to observe this line was configured 
with a total bandwidth of 234375.0 kHz and was divided into 3840 channels, providing a 
spectral channel resolution of 61.035 kHz or 0.15 km s$^{-1}$. However, the spectral 
resolution in the data cube was smoothed to 0.25 km s$^{-1}$, in order to have a better 
sensitivity in each velocity channel. 

Observations of the quasar J0510$+$1800 provided the absolute scale for the flux density
calibration, while the quasar J0431$+$2037 provided the gain calibration. 
The quasar J0510$+$1800 was also used for the bandpass calibration.

The data were calibrated, imaged, and analyzed using the Common
Astronomy Software Applications (CASA), see \citet{par2015,pare2015}. To improve the surface
brightness sensitivity, the $^{12}$CO(1-0) data were tapered to a resolution of about 1.0$''$.
The resulting r.m.s.\ noise for the line emission was about 10 mJy beam$^{-1}$ per velocity channel at an angular
resolution 0.98$''$ $\times$ 0.90$''$ with a P.A. of $-$8.66$^\circ$. We used a robust parameter of 0.5 in the 
CLEAN task, in order to obtain a better signal-to-noise ratio, at the cost of losing some spatial resolution. 

The HST/ACS (narrow band FR656N) image from XZ Tau was obtained from the HST archive 
and this is only used as a reference for the ALMA data. This HST image has been extensively 
discussed in \citet{kri2008}.    

\section{Results and Discussion}
    
In Figure 1 (right panel), we present the resulting 3 mm continuum
ALMA image of the XZ Tau. The positions of both the XZ Tau A and XZ Tau B components are shown in the left panel of
Figure 1, superposed on an optical HST/ACS (narrow band FR656N) image of the eruptive outflow \citep{kri2008}. This figure reveals 
the millimeter thermal emission arising from the binary system (A/B) located in XZ Tau and locates the continuum sources with respect 
to the outflow.  The continuum emission at 3.3 mm
is presented and discussed in \citet{par2015}. 

\begin{figure}[ht]
\begin{center}
\includegraphics[scale=0.36]{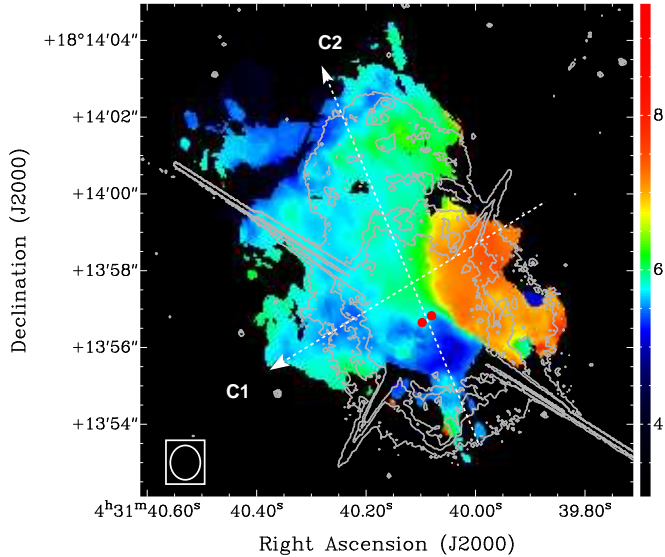}
\caption{ ALMA first moment map (or the intensity weighted velocity map) of the CO(1-0) line emission 
from the bubble-like outflow in XZ Tau overlaid in grey 
contours with an HST/ACS optical image (using the spectral filter FR656N). We have marked with red dots the positions of the ALMA  sources. 
The grey contours range from 0.05\% to 0.2\% in steps of 0.06\%  of the peak of the emission. The synthesized beam of the ALMA  
CO(1-0) image is shown in the lower-left corner. 
The optical image has an offset of about 0.4$''$ to the south in order to correct for the proper motions since the epoch of the HST observations.
The dashed arrows mark the position of the cuts (C1/C2) and their orientations toward the positive values (see Fig. 3).}
\label{fig2}
\end{center}
\end{figure}   

 
We present the first moment (that is, the velocity-weighted intensity) map of the CO(1-0) line emission in Figure 2 (the velocity range of integration is from $-$2.25 to $+$11.75 $\kms$).
  This image reveals the gas kinematics of the eruptive outflow associated with XZ Tau. We have additionally overlaid the CO velocity field on the HST/ACS optical image 
(using the spectral filter FR656N) and  we have indicated with points the positions of the ALMA 3.3 mm continuum sources. 
This image shows the structure of the CO gas 
in the entire outflow and in the vicinity of the optical jet.  These observations therefore reveal, for the first time, the kinematics 
of the molecular flow. 
The systemic LSR velocity of the cloud where XZ Tau is located is $\sim$6.8 km s$^{-1}$ (as derived from the $^{13}$CO observations of Calvet et al, 1983,
see also Partnership et al. 2015a). The velocity centroid of the expanding structure is $\sim$5.8 km  s$^{-1}$ (see Fig. 3). 
Then, as a whole, this structure is slightly blueshifted  (by about 1 km s$^{-1}$) with respect to the ambient gas. 
There is a clear velocity gradient with a southeast-northwest orientation close
to the position of XZ Tau. This molecular gas could be 
associated with expanding gas observed in the hot optical bubbles revealed by the HST at optical wavelengths. 
The CO gas associated with the older eruptions (located far from the XZ Tau multiple system) of the flow is mostly associated with 
low velocity gas close to the ambient material, presumably because it is decelerating.

\begin{figure}[ht]
\begin{center}
\includegraphics[scale=0.44]{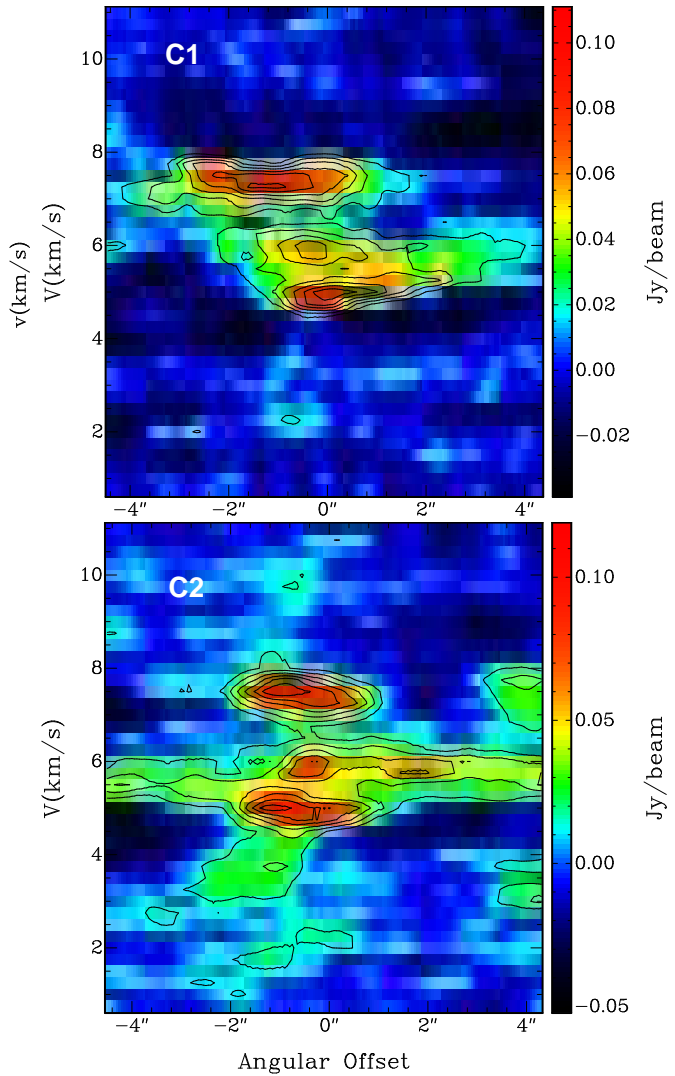}
\caption{Position$-$velocity diagrams across (P.A. $=$ 124$^\circ$) and along (P.A. $=$ 24$^\circ$) the molecular outflow in XZ Tau: 
radial velocity as function of on-the-sky distance. In Figure 2, we show the position of both cuts (C1 and C2). The angular offsets are in arcseconds 
and the LSR radial velocities in km s$^{-1}$. In both panels the contours  range from 40\% to 90\% of the peak emission, in steps of 10\%. 
The peak of the line emission is 0.11 Jy beam$^{-1}$. The synthesized beam is 0.98$''$ $\times$ 0.90$''$ with a P.A. 
of $-$8.66$^\circ$, and the spectral resolution is 0.25 km s$^{-1}$. }
\label{fig3}
\end{center}
\end{figure}

\begin{figure*}[ht]
\begin{center}
\includegraphics[scale=0.49]{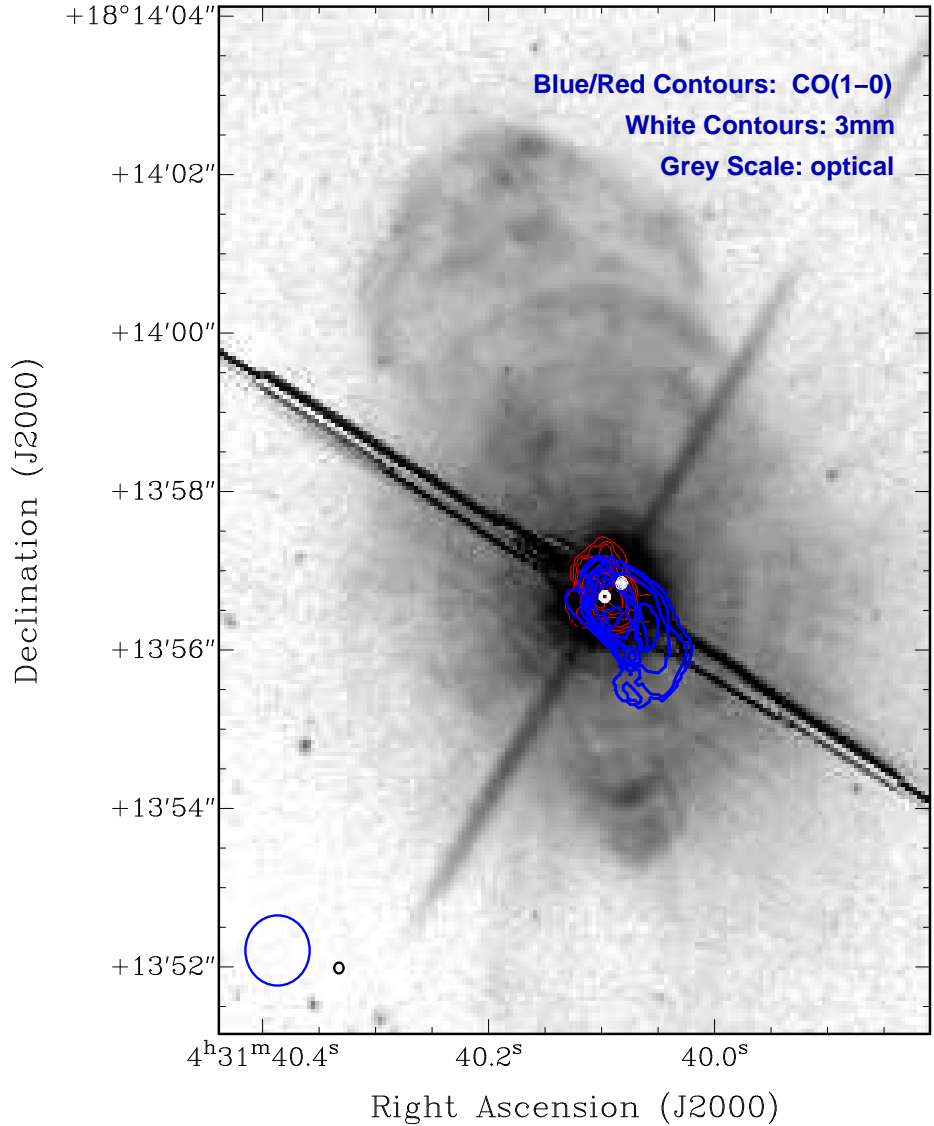}
\caption{ALMA high-velocity integrated CO(1$-$0) emission (zero moment map) overlaid on an HST/ACS optical image (narrow band FR656N) 
from the XZ Tau region. The white contours mark the 3 mm continuum emission from the binary in XZ Tau. The blue contours represent molecular emission that is 
approaching to us, while the red contours represent the emission that is receding. The integrated range of velocities for the blue contours is from 
$+$0.75 to $+$4.5 km s$^{-1}$ and for the red emission is from $+$8.25 to $+$10.75 km s$^{-1}$.  
The white contours are in percent of the peak emission starting from 10\% to 90\% in steps of 10\%. The intensity peak emission is 1.8 mJy beam$^{-1}$.
The blue and red contours are similar, starting from 30\% to 90\% in steps of 10\% for the blue contours and 15\% to 45\% in steps of 5\% for the red ones.
The peaks of the emission are 0.052 (blue) and 0.028 (red) Jy beam$^{-1}$ km s$^{-1}$.
The synthesized beams of the ALMA  CO(1-0) (blue contour) and 3 mm continuum (black contour) images are shown in the lower-left corner. 
The optical image has an offset of about 0.4$''$ to the south in order to correct for the proper motions since the HST epoch, and the WCS uncertainties 
for the HST images.}
\label{fig4}
\end{center}
\end{figure*}       
      
Figure 3 shows the position-velocity (PV) diagrams for two cuts (C1 and C2) across (P.A. $=$ 124$^\circ$) and along (P.A. $=$ 24$^\circ$) the eruptive outflow (Figures 2 and 3), which reveal the kinematics of the molecular gas with respect to the position.  The intersection point of these position-velocity diagrams is placed at the estimated centroid of the low-velocity emission, about 1 arcsec to the north of the stellar system.
These position-velocity cuts reveal an expanding and rotating molecular structure. 
The expansion is evident in the velocity components at about $+$5 and $+$7 km s$^{-1}$ 
and the rotation in the (south)east-(north)west velocity gradient.  Deeper observations are needed to 
refine our kinematic interpretation for this rotating and expanding structure. 

Three more significant features in the PV cuts 
are present: a central feature at V $\simeq$ 6 km s$^{-1}$ and 0 positional offset with respect to the intersection of the cuts, extended 
emission close to V$_{sys}$ (most notable in the C2 cut), and fainter higher velocity blueshifted and redshifted emission mainly along the C2 cut. 
This higher velocity is discussed below.


Finally, in Figure 4 additionally to the 3 mm continuum and the optical HST images, we overlaid the integrated intensity (moment 0) to the higher velocity map 
of the $^{12}$CO(1-0) emission. In the image, the blue and red colors correspond to the moderately-high velocity blueshifted and redshifted gas emission, respectively. 
We integrated  LSR velocities from $+$0.75 to $+$4.5 km s$^{-1}$  for the blueshifted lobe, and from $+$8.25 to $+$10.75 km s$^{-1}$ for the redshifted one.  This image reveals a compact bipolar molecular outflow emanating from XZ Tau. This outflow is narrow and collimated, similar in morphology to a high-velocity molecular  outflow \citep{zap2005}. However, the molecular gas seen at these moderate velocities is likely not the original gas ejected from the star/disk system but mostly entrained ambient material. The jet shows a wavy morphology, observed mostly in the blueshifted lobe (see Figure 4), possibly caused by a binary associated with component A \citep{car2009}. The P.A. of the jet is approximately  20$^\circ$. 

Morphologically, the blueshifted lobe seems to connect with one optical knot 
far from the origin. The redshifted side is more compact. The molecular outflow is ejected
from the XZ Tau A object, as first suggested by \citet{kri2008} using HST optical observations. 
These ALMA data confirm the suggestion of Krist et al. that the origin of the flow is XZ Tau A. It is interesting
to mention that the orientations of the CO outflow here reported and the optical jet/outflow as
seen in [SII] and [NII] lines (Mundt et al. 1990, Hirth et al. 1997) agree, but the blue and
redshifted sides appear to be inverted. This could suggest that they are independent features
in the XZ Tau system. Another possible explanation for this discrepancy is that while the optical jet traces 
fast, collimated, and redshifted atomic gas, the CO is tracing part of an entrained molecular flow that, 
if wide enough, could have a blueshifted component 	\citep{taka2011}.

 
 Assuming that the $^{12}$CO(1-0) line emission is optically thin and in Local Thermodynamic Equilibrium, we estimate the outflow mass using the following equation \citep[e.g.,][]{sco1986,palau2007}, which for the (1-0) transition is given by:

\small
$$
\left[\frac{M_{H_2}}{M_\odot}\right]=3.04 \times10^{-15}\,T_\mathrm{ex}\,e^{\frac{5.53}{T_\mathrm{ex}}}
\,X_\frac{H_2}{CO} 
\left[\frac{\int \mathrm{I_\nu dv}}{\mathrm{Jy\,km\,s}^{-1}}\right]
\left[\frac{\theta_\mathrm{maj}\,\theta_\mathrm{min}}{\mathrm{arcsec}^2}\right]
\left[\frac{D}{\mathrm{pc}}\right]^2, 
$$ 
\normalsize

\noindent where we used 2.8 as mean molecular weight, and X$_\frac{H_2}{CO}$  is the abundance ratio between the molecular hydrogen and the carbon monoxide ($\sim$$10^4$, e.g., Scoville et al. 1986), $T_\mathrm{ex}$ is in units of K and assumed to be 10~K, $\int \mathrm{I_\nu dv}$ is the average intensity integrated over velocity, $\theta_\mathrm{maj}$ and $\theta_\mathrm{min}$ are the projected major and minor axes of the outflow lobe, and $D$ is the distance to the source (150~pc). We then estimate a mass for the molecular collimated outflow powered by XZ Tau A of approximately 3 $\times$ $10^{-7}$ M$_\odot$.


This value is quite low compared to 
the molecular mass estimated for outflows excited by young low-mass protostars and observed with interferometers: HH 211, \citet{palau2006}; HL Tau  \citet{lum2014}; B5-IRS1 \citep{zap2014}; and DG Tau B  \citep{zap2015}, and even lower, but comparable, to those observed in molecular outflows from sub-stellar objects \citep[e.g.][]{phan2014}. 
We also estimated a kinetic energy of 3 $\times$ 10$^{37}$ ergs. 

Taking a radial velocity for the CO of 5 $\kms$ (which are the velocities for the moderately-high velocity collimated outflow), a size of 1$''$ (see Figure 4) 
and an angle with respect to the plane of the sky of 30$^\circ$, 
we estimate a dynamical age of about  t$_{dyn}$ = (size of the outflow in the plane of the sky / radial 
velocity) $\times$ $\tan [\theta]$ = (2.3$\times$10$^{15}$ cm / 5 $\times$10$^5$ cm s$^{-1}$) $\times$ $\tan [\theta]$  = 148  $\tan [\theta]$  years for 
the collimated molecular outflow. In this equation $\theta$ is
the angle of the outflow with respect to the plane of the sky.   
Using the determinations and respective errors for the radial and transversal velocities 
\citep{hir1997,kri2008} for
the optical jet, we find that $\theta$ can be in a wide range of values,
from 13 to 61 degrees. This implies an age estimate between 33 and 266 years,
with the smaller value close to the estimated  age of 34 years for the bubble
by the time the ALMA data was taken. Then, the ejection of the optical bubbles and
the creation of the molecular collimated outflow could have been simultaneous, however the fact that the
orientations are inverted (for the collimated outflow and the bubbles) suggests they are probably different 
structures.

\section{Conclusions}

We have analyzed sensitive CO line ALMA observations in the millimeter regime from the young multiple star XZ Tau.  
Our conclusions are as follows: 
 
\begin{itemize} 

\item  The $^{12}$CO(1-0) line observations show for the first time the kinematics of the molecular outflow.  
The kinematics of the different ejections close to XZ Tau reveal a slow ($\Delta$V $=$ 2 km s$^{-1}$) and expanding/rotating 
 structure. This structure appears to be related with the multiple eruptive bubbles unveiled in the optical HST images. 

\item  The observations also reveal the presence 
of a compact and collimated bipolar molecular outflow (reminiscent of a jet)  with a 
northeast-southwest orientation (P.A. $\sim$ 20$^\circ$) emanating from XZ Tau A.  
This outflow is probably tracing the base of the optical eruptive flow, with a similar orientation.
The mass (3 $\times$ 10$^{-7}$ $\solarmass$) and energetics (3 $\times$ 10$^{37}$ ergs) for the molecular 
collimated outflow are smaller than those values found in other molecular collimated outflows associated with 
young low-mass stars.

\item The observations confirm the origin of the eruptive outflow as XZ Tau A. These observations argue 
in favor of the model presented by \citet{car2009} with regard to the 
periodic nature of the outflow, however, as discussed in \citet{for2014}, additional observations at different wavelengths 
are needed, especially dedicated observations to follow up the motions, and the outflow ejections from the components 
in the XZ Tau multiple star system.  This will help to identify firmly the physical mechanisms behind the ejection of the bubble-like flow.     

\end{itemize}  
 
\acknowledgments

All the authors of this paper acknowledge the financial support from DGAPA, UNAM, and CONACyT, M\'exico. 
A.P. acknowledges the financial support from UNAM-DGAPA-PAPIIT IA102815, M\'exico.
This paper makes use of the following ALMA data: ADS/JAO.ALMA$\#$2011.0.00015.SV. 
ALMA is a partnership of ESO (representing its member states), NSF (USA) and 
NINS (Japan), together with NRC (Canada) and NSC and ASIAA (Taiwan), 
in cooperation with the Republic of Chile. 
The Joint ALMA Observatory is 
operated by ESO, AUI/NRAO and NAOJ. The National Radio Astronomy Observatory
is a facility of the National Science Foundation operated under cooperative 
agreement by Associated Universities, Inc.
Based on observations made with the NASA/ESA Hubble Space Telescope, 
obtained from the data archive at the Space Telescope Science Institute. STScI is operated by the Association 
of Universities for Research in Astronomy, Inc. under NASA contract NAS 5-26555.
We are very thankful for the suggestions of the anonymous referee that helped to improve our manuscript.

\end{document}